\documentclass[aps,prl,twocolumn]{revtex4-1}

\include{setspace}
\include{dcolumn}
\usepackage{graphicx}
\usepackage{longtable}
\usepackage{multirow}
\begin{document}

\title{Testing the Mutually Enhanced Magicity Effect in Nuclear Incompressibility via the Giant Monopole Resonance in the $^{204,206,208}$Pb Isotopes}

\author{D. Patel$^{1,2}$,\, U. Garg$^{1,2}$,\, M. Fujiwara$^{3}$,\, T. Adachi$^{4}$,\, H. Akimune$^{5}$,\, G.P.A. Berg$^{1,2}$,\, M.N. Harakeh$^{4,6}$,\, M. Itoh$^{7}$,\,\\ C. Iwamoto$^{5}$,\, A. Long$^{1,2}$,\, J.T. Matta$^{1,2}$,\, T. Murakami$^{8}$,\, A. Okamoto$^{5}$,\, K. Sault$^{1}$,\, R. Talwar$^{1,2}$,\, M. Uchida$^{9}$,\, M. Yosoi$^{3}$}

\affiliation{$^1$Department of Physics, University of Notre Dame, Notre Dame, Indiana 46556, USA\\ $^2$Joint Institute for Nuclear Astrophysics, University of Notre Dame, Notre Dame, Indiana 46556, USA\\
$^{3}$Research Center for Nuclear Physics, Osaka University, Osaka 567-0047, Japan\\$^{4}$Kernfysisch Versneller Instituut, University of Groningen, 9747 AA Groningen, The Netherlands\\ $^5$Department of Physics, Konan University, Kobe 568-8501, Japan\\ $^6$GANIL, CEA/DSM-CNRS/IN2P3, 14076 Cean, France\\$^{7}$Cyclotron and Radioisotope Center, Tohoku University, Sendai 980-8578, Japan\\$^{8}$Division of Physics and Astronomy, Kyoto University, Kyoto 606-8502, Japan\\$^{9}$Department of Physics, Tokyo Institute of Technology, Tokyo 152-8850, Japan}

\date{\today}

\begin{abstract}
Using inelastic $\alpha$-scattering at extremely forward angles, including $0^\circ$, the strength distributions of the isoscalar giant monopole resonance (ISGMR) have been measured in the $^{204,206,208}$Pb isotopes
in order to examine the proposed mutually enhanced magicity (MEM) effect on the nuclear incompressibility. The MEM effect had been suggested as a likely explanation of the ``softness'' of nuclear incompressibility observed in the ISGMR measurements in the Sn and Cd isotopes. Our experimental results rule out any manifestation of the MEM effect in nuclear incompressibility and leave the question of the softness of the open-shell nuclei unresolved still.
\end{abstract}

\maketitle

Nuclear incompressibility, a measure of the curvature of the equation of state (EOS) at saturation density, is an important parameter in describing the shape of EOS accurately \cite{BohrMott1975}. EOS, in turn, plays an important role in describing the collective properties of nuclei as well as
the dynamical properties of astrophysical objects \cite{Prakash2001,HoroPiekare2001}.The cleanest experimental tool for measuring nuclear incompressibility is provided by the compression modes of nuclear vibration: the isoscalar giant monopole resonance (ISGMR) and the isoscalar giant dipole resonance (ISGDR). The centroid energy of the ISGMR, $E_{ISGMR}$, is directly related to the incompressibility of the finite nuclear matter, $K_A$, as:
\begin{equation}\label{eq:2}E_{ISGMR}=\hbar\sqrt{\frac{K_A}{m<r^2>}}
\end{equation}
\noindent
where, $m$ is the nucleon mass and $<r^2>$ is the mean square radius of the nucleus \cite{Strin82,trein81}.

In order to extract the nuclear matter incompressibility, $K_{\infty}$, from the $E_{ISGMR}$, one builds a class of energy functionals, $E(\rho)$, with different parameters that allow calculations for nuclear matter and finite nuclei in the same theoretical framework. The parameter-set for a given class of energy functionals is characterized by a specific value of $K_{\infty}$. The ISGMR strength distributions are obtained for different
energy functionals in a self-consistent RPA calculation. The $K_{\infty}$ associated with the interaction that best reproduces the ISGMR centroid energies is, then, considered the correct value \cite{blaizot,Tao2007}.

Following this procedure, both relativistic and non-relativistic calculations give $K_{\infty}$=240$\pm$20 MeV \cite{colo1,jorge1,shlomo}.
These accurately calibrated relativistic and non-relativistic models reproduce very well the ISGMR centroid energies in the ``standard" nuclei, $^{90}$Zr and $^{208}$Pb. However, the same models overestimate the ISGMR centroid energies in the Sn isotopes by as much as 1 MeV \cite{Tao2007,Tao2010} and, consequently, a lower value of nuclear incompressibility is required to reproduce ISGMR centroid energies in the Sn isotopes. This has led to the puzzling, and still open, question: Why are the Sn isotopes so soft? \cite{Piekare2007,garg2007}. This effect was confirmed recently in measurements on the nearby isotopic chain of $^{106,110-116}$Cd nuclei~\cite{DP2012}.

One of the first attempts to resolve this issue entailed the inclusion of the pairing interactions. However, that reduced the discrepancy by only about 150 keV in the case of Sn isotopes~\cite{cita,colo2,Jun2008,Khan-R}, and despite concerted theoretical effort in recent years~\cite{colo2,Jun2008,Khan-R,cente09,dario,kvasil,jacek}, the challenge has remained to simultaneously describe the ISGMR in the open-shell nuclei as well as in the doubly magic $^{208}$Pb and $^{90}$Zr nuclei. Indeed, this has been identified as one of the ``open'' problems in nuclear structure in a recent major compilation~\cite{major}.

An intriguing proposal put forward to explain this discrepancy has been that the mutually enhanced magicity (MEM) effect may play a role in the nuclear incompressibility \cite{Khan2009}. MEM refers to a strong underbinding observed in the Hartree-Fock mass formulas (HFMF) for all doubly magic nuclei and their immediate neighbors formed by adding or removing not more than one nucleon~\cite{Zeldes1983,Lunney2003}. As noted in Ref.~\cite{Lunney2003}, there are 27 such nuclei for which the mean error between the experimentally measured values of the nuclear masses and those calculated by HFMF is $\approx$1.31 MeV as against $\approx$40 keV for the complete set of $\textgreater$1700 data points available at the time of that compilation.

It has been argued that $^{208}$Pb, being a doubly-magic nucleus, is ``stiffer'' than the open-shell nuclei and the incompressibility obtained from doubly-magic nuclei would invariably lead to an overestimation of the ISGMR energies in the open-shell nuclei~\cite{Khan2009}. An important prediction of the inclusion of the MEM effect in the calculations of the ISGMR centroid energies in the constrained Hartree-Fock-Bogoliubov (CHFB) framework
is that the ISGMR centroid energy in $^{208}$Pb would be higher than the corresponding values
in the $^{204,206}$Pb isotopes by $\sim$600 keV. This predicted excitation-energy difference is large enough to be examined experimentally, considering the current experimental uncertainties of ISGMR measurements.

With an aim to explore the role of MEM effect in the nuclear incompressibility, and thereby attempting to address the important question of the softness of the open-shell nuclei, we have measured the ISGMR strength distributions in the $^{204,206,208}$Pb isotopes in the same experiment. We find that the ISGMR centroid energies in these three Pb isotopes are very close, in contradiction to the aforementioned theoretical prediction, and rule out the effect of MEM in nuclear incompressibility.

The experiment was performed at the ring cyclotron facility of the Research Center for Nuclear Physics (RCNP), Osaka University, Japan. Details of the experimental method and analysis procedures have been provided previously \cite{Itoh2003} and are described only briefly here. The inelastic scattering of 100 MeV/u $\alpha$ particles off the three self-supporting $^{204,206,208}$Pb targets was studied in the same experiment in order to minimize any systematic errors. Highly-enriched targets ($99.94\%$, $99.76\%$, $99.70\%$, for $^{204}$Pb, $^{206}$Pb and $^{208}$Pb, respectively) were used, with thicknesses in the range of 4.92 mg/cm$^2$ to 5.82 mg/cm$^2$. Elastic scattering measurements in the angular range of $3.5^\circ$ to $24^\circ$ were performed for $^{204}$Pb and $^{206}$Pb to obtain the optical model parameters (OMP). For $^{208}$Pb, the elastic scattering measurements were performed over $3.5^\circ$--$8.5^\circ$, and previous data from Ref. \cite{UchidaThesis} were employed for angles up to $25^\circ$. The measurements of inelastically scattered $\alpha$ particles were made over the range $0^\circ$-- $9.5^\circ$. The importance of making measurements at such extreme forward angles, including $0^\circ$, is twofold: the cross section for the ISGMR, which is the focus of this study, peaks at $0^\circ$, and the $L$=0 angular distribution is most distinct at the very forward angles. These measurements are, however, extremely difficult since the primary beam passes very close to the scattered particles at these angles and one requires a combination of a high-quality, halo-free beam, and an appropriate magnetic spectrometer; the high-resolution magnetic spectrometer Grand Raiden, at RCNP \cite{Fujiwara99} is a most suitable such instrument.

\begin{figure}
\includegraphics[width=0.49\textwidth]{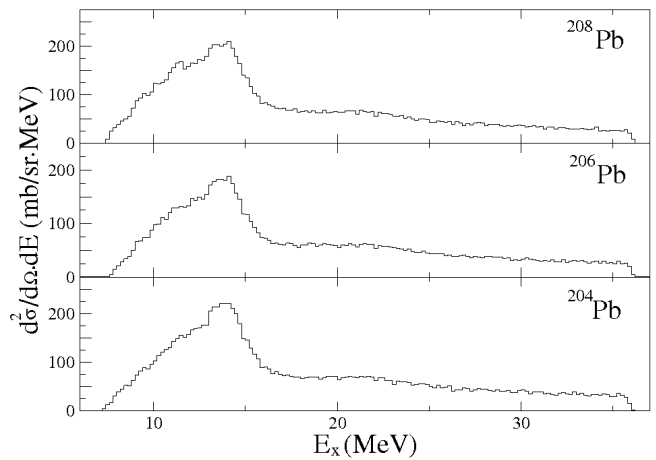}
\caption{Excitation-energy spectra at an ``average'' angle of 0.7$^\circ$ for the Pb isotopes investigated in this work.
\label{fig:0-deg}}
\end{figure}

The scattered particles were momentum analyzed by Grand Raiden and focused onto the focal plane detector system comprising of two multi-wire drift chambers (MWDC) and two plastic scintillation counters~\cite{Itoh2003}. For the 0$^{\circ}$ measurements, the Faraday cup was located 3 m downstream of the focal plane and for the other angles, Faraday cups were located in the target chamber and behind the Q1 magnet of the spectrometer~\cite{itoh-rcnp}. The focal-plane detectors allowed for the determination of the position and the incident angle of the scattered $\alpha$ particle in the focal plane, and covered the excitation energy range of $E_x$ $\simeq$ 8.5--35 MeV. Using the ray-tracing technique for determining the trajectories of the scattered particles, excitation-energy spectra were obtained for specific angles by subdividing the full angular opening. Grand Raiden was used in the double-focusing mode in order to eliminate the instrumental background~\cite{Tao2010,Itoh2003}.
The background-subtracted inelastic scattering spectra for the three Pb isotopes at $0.7^\circ$ are shown in Fig.~\ref{fig:0-deg}. The cross-section at the low excitation energy end is mainly accounted for by the isoscalar giant quadrupole resonance.

\begin{table*}[!hbt]
 \caption{\label{hn} Lorentzian fit parameters for the ISGMR strength distributions obtained for the three Pb isotopes. For comparison the ISGMR centroid energy values from the previous experiments are presented. The moment ratio, $\sqrt{m_1/m_{-1}}$, values are calculated over the energy range of 9.5-19.5 MeV. Also quoted are the values of the moment ratios obtained from the calculation including pairing and MEM effect for the three Pb isotopes.}
\begin{tabular}{|c|c|c|c|c|c|c|c|c|c|}
 \hline
 \cline{1-10}
  Target & \multicolumn{4}{|c|}{E$_{ISGMR}$ (MeV)} & $\Gamma_{ISGMR} (MeV)$ &\multicolumn{2}{|c|}{ $\sqrt{m_1/m_{-1}}$ (MeV)} & \multicolumn{2}{|c|}{ E$_{ISGMR}$A$^{1/3}$ (MeV)}\\ \cline{1-10}
   & This work & RCNP-U\footnotemark[1]& Texas A$\&$M\footnotemark[2]& KVI\footnotemark[3] & This work & This work & Pairing+MEM\footnotemark[4]  & This work & Pairing+MEM\footnotemark[4]  \\ \cline{2-5} \cline{7-10}
   \hline
  $^{204}$Pb & 13.8$\pm$0.1 & - & - & - & 3.3$\pm$0.2 & 13.7$\pm$0.1 & 13.4 & 81.2$\pm$0.6 & 78.9\\

  $^{206}$Pb & 13.8$\pm$0.1 & - & - & 14.0 $\pm$0.3 & 2.8$\pm$0.2 & 13.6$\pm$0.1 & 13.4 & 81.5$\pm$0.6 &  79.1\\

  $^{208}$Pb & 13.7$\pm$0.1 & 13.5$\pm$0.2 & 13.91$\pm$0.11 & 13.9$\pm$0.3 & 3.3$\pm$0.2 & 13.5$\pm$0.1 & 14.0 & 81.2$\pm$0.6 & 82.9\\
  \hline
\end{tabular}
\footnotetext[1] {Ref.~\cite{Uchida2003}}
\footnotetext[2] {Ref.~\cite{Youngblood1999}}
\footnotetext[3] {Ref.~\cite{Harakeh1979}}
\footnotetext[4] {Ref.~\cite{Khan2009}}
\end{table*}

Elastic scattering cross sections were used to extract the optical model parameters using the ``hybrid'' potential proposed by Satchler and Khoa \cite{khoa}. In this procedure, the density-dependent single-folding model with a Gaussian $\alpha$-nucleon potential was used to determine the real part of the optical potential. The computer codes SDOLFIN and DOLFIN \cite{RickDOLF} were used to calculate the shape of the real part of the potential and the form factors, respectively. For the imaginary term, a Woods-Saxon potential was used and its parameters ($W_I$, $r_I$, and $a_I$), together with the depth of the real part, $V_R$, were obtained by fitting the elastic-scattering cross sections using the $\chi^2$ minimization technique, with the help of the computer code PTOLEMY \cite{Brown80}. Using the established $B(E2)$ and $B(E3)$ values \cite{nndc} and the OMP thus obtained, the angular distributions of the differential cross sections for the $2^+_1$ states in $^{204,206}$Pb and the $3^-_1$ state in  $^{208}$Pb were calculated. A good agreement between the calculated and the experimental angular distributions of differential cross sections of these states established the validity and appropriateness of the extracted OMP.

The inelastic-scattering cross sections were binned with a 1-MeV width to reduce statistical fluctuations. In order to extract the ISGMR contribution from the excitation-energy spectra at different scattering angles, multipole decomposition analysis (MDA) was performed whereby the experimental differential cross sections were expressed as linear combinations of the calculated distorted-wave Born approximation (DWBA) differential cross sections:

\begin{equation}
\label{eq:4}
\frac{d^2\sigma^{exp}(\theta_{c.m.},E_x)}{d\Omega dE}=\sum_{L}a_{L}(E_x)\frac{d^2\sigma^{DWBA}_L(\theta_{c.m.},E_x)}{d\Omega dE}
\end{equation}

\noindent
where, $L$ is the order of multipole.
The transition densities and sum rules for the various multipoles used in these calculations are discussed in Refs.~\cite{HarakehBook, Satchler87}. DWBA calculations were performed for up to a maximum angular momentum transfer of $\triangle L =7$. Photonuclear data were used in conjunction with DWBA calculations based on the Goldhaber-Teller model to estimate the contribution of the isovector giant dipole resonance (IVGDR)~\cite{Data}. No experimental photonuclear cross-section data are available for $^{204}$Pb; in this case, the parameters were obtained from a global equation derived from the best fit to the available data over a wide range of nuclear masses \cite{Berman1975}. In the event, the corresponding IVGDR cross sections are too small to affect the MDA results for the ISGMR in any significant way. Further details of the analysis procedures, as well as all the MDA fits will be provided elsewhere~\cite{DP.thesis}.

\begin{figure}
\includegraphics[width=0.49\textwidth]{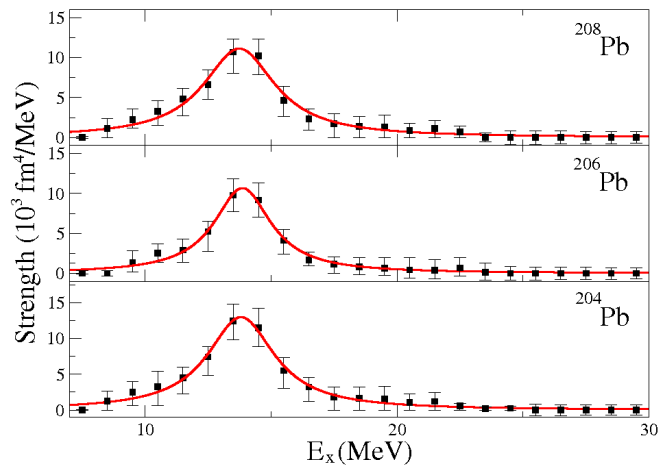}
\caption{The ISGMR strength distributions of the Pb isotopes investigated in this work. Solid lines represent Lorentzian fits to the data.
\label{fig:strength}}
\end{figure}

The ISGMR strength distributions extracted from the MDA for the three Pb isotopes investigated in this work are plotted in Fig.~\ref{fig:strength}. The centroid energies obtained from the Lorentzian fits to the distributions (also shown in Fig.~\ref{fig:strength}), along with the moment ratios $\sqrt{m_1/m_{-1}}$ for the strength distributions, are listed in Table I. The moment ratios were computed over the energy range 9.5--19.5 MeV using the definition $m_k=\int \! E^k\ \! S(E)\, \mathrm{d} E$; this energy range incorporates nearly all of the ISGMR strength. As is clear from the data presented in Table I, the experimental ISGMR centroid energies are very close to each other for the three Pb isotopes, in stark disagreement with the predicted $\sim$600 keV difference in the ISGMR centroid energies of $^{204}$Pb and $^{208}$Pb resulting from the MEM effect. Indeed, the product $E_{ISGMR}A^{1/3}$ is practically identical for the three isotopes (see Table I), indicating that the ISGMR centroid energies follow the standard A$^{-1/3}$ dependence. These results are in contrast with the theoretical predictions of Ref. \cite{Khan2009}, where this quantity varies by 4 MeV between $^{204}$Pb and $^{208}$Pb. We further note that the difference in $E_{ISGMR}$ for $^{204}$Pb and $^{208}$Pb due to the $K_{\tau}$ value obtained from the measurements on the Sn and Cd isotopes \cite{Tao2007,Tao2010,DP2012} would be 0.19$\pm$0.03 MeV, consistent with the current measurements. 


To summarize, we have measured the ISGMR strength distributions in the $^{204,206,208}$Pb isotopes in the same experiment in an attempt to test the predicted manifestation of the MEM effect on the ISGMR centroid energies in open-shell nuclei. A careful extraction of the ISGMR strength distribution shows that the ISGMR energies are practically identical in the three isotopes, ruling out any consequences of the MEM effect in nuclear incompressibility. The MEM effect, thus, does not account for the observed softness of the Sn and the Cd isotopes, and this question remains open still.

We acknowledge the effort of the RCNP staff in providing high-quality $\alpha$ beams required for the present measurements. This work has been supported in part by the National Science Foundation (Grant Nos. PHY07-58100 and PHY-1068192).

\end{document}